\documentclass[prb,twocolumn,aps]{revtex4}
\usepackage{graphicx}% Include figure files
\usepackage{dcolumn}% Align table columns on decimal point
\usepackage{bm}% bold math
\usepackage{amssymb}
\usepackage{amsmath,amsfonts}
\renewcommand{\vec}[1]{{\mathbf{#1}}}
\newcommand{\beq}{\begin{eqnarray}}
\newcommand{\eeq}{\end{eqnarray}}
\newcommand{\q}{\quad}

\begin{document}

\title{Classification of quantum phases for the star-lattice antiferromagnet
via a projective symmetry group analysis}

\author{Ting-Pong Choy}
\affiliation{Department of Physics, University of Toronto, Toronto, Ontario M5S 1A7, Canada}

\author{Yong Baek Kim}
\affiliation{Department of Physics, University of Toronto, Toronto, Ontario M5S 1A7, Canada}

\date{\today}

\begin{abstract}
We study possible quantum ground states of the Heisenberg antiferromagnet on the star lattice, which may be
realized in the recently discovered polymeric Iron Acetate, Fe$_3$($\mu_3$-O)($\mu$-OAc)$_6$(H$_2$O)$_3$[Fe$_3$($\mu_3$-O)($\mu$-OAc)$_{7.5}$]$_2\cdot$ 7H$_2$O.\cite{zheng}
Even though the Fe$^{\rm III}$ moment in this material carries spin-5/2 and the system eventually orders magnetically at low temperatures, the magnetic
ordering temperature is much lower than the estimated Curie-Weiss temperature, revealing the frustrated nature of the spin interactions.
Anticipating that a lower spin analog of this material may be synthesized in future, we investigate the effect of quantum fluctuations on the
star-lattice antiferromagnet using a large-$N$ Sp($N$) mean field theory and a projective symmetry group analysis for possible bosonic quantum spin liquid phases. It is found that there exist only two distinct gapped $Z_2$ spin liquid phases with bosonic spinons for non-vanishing nearest-neighbor valence-bond-amplitudes. In particular, the spin liquid phase which has a lower energy in the nearest-neighbor exchange model can be stabilized for relatively higher spin magnitudes. Hence it is perhaps a better candidate for the realization of quantum spin liquid state. We also determine the magnetic ordering patterns resulting from the condensation of the bosonic spinons in the two different spin liquid phases. We expect these magnetic ordering patterns would directly be relevant for the low temperature ordered phase of the Iron Acetate. The phase diagram containing all of these phases and various dimerized states are obtained for the nearest-neighbor exchange model and its implications are discussed.

%Frustrated star lattice is a generalization of Kagome lattice by introducing an additional link between any two adjacent corner-sharing triangles. It is first realized in Fe$_3$($\mu_3$-O)($\mu$-OAc)$_6$(H$_2$O)$_3$[Fe$_3$($\mu_3$-O)($\mu$-OAc)$_{7.5}$]$_2\cdot$ 7H$_2$ O by Zheng et. al.\cite{zheng}. In this paper, we present the phase diagram of frustrated star lattice using Sp(N) theory. Distinct phases including dimerized states, quantum paramagnets and spin-ordered states are obtained in the large-N limit. We classified different quantum spin liquids using projective-symmetry group (PSG) method.  By assuming non-vanishing nearest-neighbor amplitudes, there are only two distinct $Z_2$ spin liquids which carrying zero and $\pi$ flux respectively. The $\pi$-flux state has a significantly large values of quantum parameter $\kappa_c$ at which magnetic order occurs. We argue that this $\pi$-flux state is an attractive candidate as the ground state of star lattice in the presence of ring-exchange coupling.

\end{abstract}

\maketitle

\section{Introduction}

\begin{figure}[t]
\centering
\includegraphics[width=7.cm]{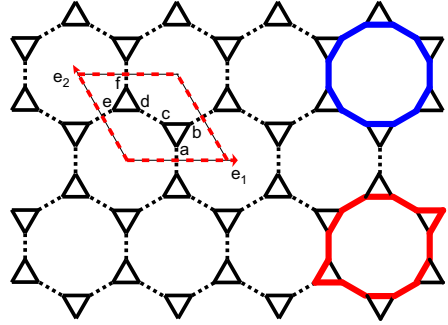}
\caption{(color online) Star lattice is shown with two inequivalent nearest neighbor spin exchange interactions $J_t$ and $J_d$ along the triangular and bridge links,
denoted by the solid and dotted lines respectively. The rhombus enclosed by dashed lines corresponds to a unit cell with six sites labeled by the indices $a$ to $f$. 
Here the 12-sided (blue) and 14-sided (red) loops are also shown.} 
\label{star}
\end{figure}

The search for quantum spin liquid phases in two and three dimensions has lead to recent discoveries of several spin-1/2 frustrated antiferromagnets, where no magnetic ordering has been seen down to the lowest temperature. The examples include a triangular lattice organic material close to a metal-insulator transition\cite{shimizu}, Kagome or Kagome-like lattice systems \cite{helton,hiroi,okamoto-Kagome}, and a three-dimensional hyber-Kagome lattice material \cite{okamoto}. The nature of possible spin liquid and other competing phases in these systems has been a subject of intense research activities. While there has been considerable progress in understanding some of the candidate quantum paramagnetic phases such as quantum spin liquid\cite{ran,ryu,lawler2008,zhou,podolsky,lecheminant,waldtmann} and valence bond solid phases\cite{nikolic,singh,yang2008}, a general understanding of the interplay between competing phases upon the variation of the spin interactions is still lacking\cite{lawler2007}. Therefore, systematic studies of a variety of frustrated magnets with possibly different spin interactions and/or with different underlying lattice structures would be extremely useful.\cite{wang2007}

In this regard, the recent discovery of the Iron Acetate may present one of such useful examples for a two-dimensional frustrated lattice\cite{zheng}. Here Fe$^{\rm III}$ spin-5/2 moments reside on the star lattice as shown in Fig.\eqref{star}. The Curie-Weiss temperature is estimated to be $\Theta_{\rm CW} = - 581$K, but the magnetic ordering occurs only below $T_N = 4.5$K (the nature of the magnetic order is presently not known), leading to a large frustration parameter, $f = |\Theta_{\rm CW}|/T_N = 129 \gg 1$.  This raises the hope that spin liquid phases may exist for a lower spin analog of this material.

In this paper, we investigate possible quantum ground states of the star-lattice antiferromagnet, including quantum spin liquid phases, magnetically ordered states, and dimerized phases using a projective symmetry group analysis\cite{q_order} and a large-$N$ Sp($N$) mean-field theory\cite{sachdev-Kagome,fawang}. We expect the quantum paramagnetic phases, namely the spin liquid and dimerized phases, may be relevant to a lower spin analog ({\it e.g.} spin-1/2 or spin-1) of the Iron Acetate, which is yet to be discovered. The magnetically ordered phases described in this work may directly be relevant to the low temperature ordered phase of the Iron Acetate.

The star lattice can be regarded as a triangular Bravais lattice with a six-site basis and hence the unit cell contains six lattice sites as shown in Fig.\eqref{star}. One can also view this lattice as a variant of the Kagome lattice in the sense that additional lattice links between triangles of the Kagome lattice are introduced. This leads to two topologically in-equivalent nearest-neighbor spin exchange interactions: $J_t$ along the {\it triangular links} and $J_d$ along the {\it bridge links} that connect triangles.
%In the case of the nearest-neighbor spin exchange model, if $J_d < 0$, every two spins connected by a bridge link would be in parallel and the system reduces to a Kagome lattice with the nearest-neighbor spin coupling, $J_t$. Similarly, when $J_t<0$, all three spins on the same triangle would favor the same direction and the system can be mapped to a honeycomb lattice with the spin coupling, $J_d$.
In the Heisenberg model with the antiferromagnetic sign for both $J_d$ and $J_t$ ($J_t > 0, J_d > 0$), there is clearly a macroscopic degeneracy of the classical ground states.\cite{richter}
%Notice that the classical energy can be minimized by making the directions of two spins connecting the triangles to be anti-parallel, hence every classical ground state of the star lattice with $N_s$ sites can be mapped to one of the Kagome lattice system with $N_s/2$ sites. Thus the degeneracy of these two systems are basically identical and macroscopically large.
Previous exact diagonalization studies of the spin-1/2 nearest-neighbor antiferromagnetic Heisenberg model on the star lattice suggest that the ground state may be a dimerized state with dimers sitting on the bridge links for $J_t=J_d$ and a 3-fold degenerate valence bond solid state when $J_t > 1.3 J_d$.\cite{misguich,richter} The finite size effect in these studies, however, makes it difficult to draw a definite conclusion. Various models including the quantum dimer model\cite{fjaerestad} and Kiteav model\cite{yao} on the star lattice have also been studied recently.

In this work, we provide systematic understanding of possible quantum spin liquid phases with bosonic spinons in the star-lattice antiferromagnet using a projective symmetry group analysis of the mean-field states in the Schwinger boson theory. The projective symmetry group is a powerful tool to classify all and only the physical spin liquid states without specifying a particular spin Hamiltonian. We also investigate how these spin liquid states may be related to the previously-identified dimerized phases\cite{misguich,richter} in the global phase diagram using a large-$N$ Sp($N$) Schwinger boson mean field theory. 
%Notice that these spin liquid states cannot be distinguished by conventional broken symmetries. The main idea of the analysis is that two apparently different mean-field states may correspond to the same physical quantum spin liquid state if they are related by a combination of certain space and gauge group operations (once they survive the appropriate projection to the physical Hilbert space). The set of the physically allowed combinations of space and gauge group operations forms a projective symmetry group. In this way, one would not miss out a quantum spin liquid state that may not be stabilized in a particular Hamiltonian, but may be a ground state in a slightly different spin model. This method has been successfully applied to the spin liquid phases with fermionic spinons on various lattices and the spin liquid phases with bosonic spinons on the Kagome lattice.

Here we focus on the $Z_2$ spin liquid phases on the star lattice, where $Z_2$ represents a global pure gauge degree of freedom that leaves the mean-field states invariant. It has been shown that such $Z_2$ spin liquid states naturally arise in the Schwinger boson theory of the antiferromagnetic Heisenberg model on frustrated lattices.\cite{sachdev-Kagome} The projective symmetry group analysis leads to only a finite number of such $Z_2$ spin liquid phases. If we further require the system to have only non-trivial nearest-neighbor valence bond amplitudes, there exist only two distinct $Z_2$ (symmetric) spin liquid phases that preserve all the space group, spin rotation, and time reversal symmetries in contrast to the four symmetric $Z_2$ spin liquid states on the Kagome lattice\cite{fawang}. These two states can also be distinguished by the ``flux" enclosed in the 12-sided loop as shown in Fig.\eqref{star}, which is defined as the phase of the gauge-invariant product of the valence-bond-amplitudes $Q_{ij}$ along the 12-sided loop, {\it i.e.} $\Theta = \arg [Q_{1,2}(- Q_{2,3}^*)Q_{3,4} \cdots (-Q_{12,1}^*)]$.\cite{oleg} The two $Z_2$ spin liquid phases correspond to $\Theta=0$ and $\Theta=\pi$, respectively. The zero-flux state is an analog of the [$0$Hex,$\pi$Rhom] phase of the Kagome lattice\cite{fawang}. We study the spinon and spin-1 excitation spectra in the two spin liquid phases. In principle, the spin-1 excitation spectra can be measured by neutron scattering experiment to distinguish these two phases when an ambiguity as to the nature of the underlying quantum paramagnetic phase arises.

Using the results above, we also investigate possible magnetically ordered phases via the condensation of bosonic spinons in each spin liquid phase. The magnetic order arising from the zero-flux state has the magnetic ordering wavevector ${\bf q} = \pm(\frac{\pi}{3}+n\pi, \frac{\pi}{3}+m\pi)$ and $(n\pi,m\pi)$ with integers $n,m$. On the other hand, the magnetic ordering arising from the $\pi$-flux state has the ordering wavevector ${\bf q} = (0,0)$ and does not break translational symmetry. These results may directly be relevant to the low temperature magnetically ordered state of the Iron Acetate. The determination of the magnetic ordering wavevector would also tell us which spin liquid phase may close by.\cite{isakov}

The relative stability of all these phases and the previously studied dimerized states\cite{misguich,richter} is studied in a large-$N$ Sp($N$) mean field theory of the nearest-neighbor exchange model\cite{sachdev-Kagome,fawang} and the global phase diagram is obtained as a function of the effective spin magnitude $\kappa=2S_{\rm eff}$ and $J_t/J_d$. The advantage of the large-$N$ Sp($N$) theory is that one can treat the magnetically ordered and paramagnetic states on equal footing and the method is non-perturbative in the effective spin magnitude, $\kappa = 2S_{\rm eff}$. The results are shown in Fig.\eqref{phase-p1=0}. It is found that the zero-flux state is always energetically favorable over the $\pi$-flux state in the nearest-neighbor model. In contrast to Kagome lattice, the critical $\kappa$ beyond which a magnetic order sets in, is much larger for the zero-flux state, {\it i.e.} $\kappa_c$ of the zero-flux phase can be as large as $\kappa_c \sim 5$ while the largest $\kappa_c \sim 1.5$ for the $\pi$-flux state. $\kappa_c \sim 5$ is an unusually large number because $\kappa_c$ is often smaller than unity in many cases.\cite{fawang} In fact, this is even larger than $\kappa_c \sim 2$ of  the [$0$Hex,$\pi$Rhom]  phase of the Kagome lattice. This suggests that the zero-flux phase may exist even for relatively large spin ($S>1/2$) system in an anisotropic limit.

In the ultimate quantum limit, $\kappa \ll 1$, the dimerized state with the dimers sitting on the
$J_d$ bonds becomes the ground state when $J_d > J_t$ while only the spin correlations on the
triangles survives in the opposite limit, $J_d < J_t$ (for the nearest-neighbor model).\cite{oleg}
The dimerized state for $J_d > J_t$ is consistent with the previous numerical result\cite{misguich,richter} on the spin-1/2 nearest-neighbor Heisenberg model. The nature of the dimerized state for $J_d < J_t$ cannot clearly be identified in the present work because it requires further analysis of the $1/N$ fluctuation
corrections.\cite{sachdev89,affleck89} We emphasize, however, that the phase boundaries of various phases may look different in the physical $N=1$ limit, so the phase diagram obtained in the large-$N$ limit should be taken with a grain of salt. Further, it is possible that the inclusion of other spin interactions may favor the spin liquid over the dimerized states even deep inside the quantum regime $\kappa \ll 1$. The nature of the transitions between various phases in the phase diagram is also discussed in the main text of the paper.

The rest of the paper is organized as follows. In Sec.~\ref{sec-meanfield}, we briefly review an Sp($N$) mean field theory of the antiferromagnetic Heisenberg model. In Sec.~\ref{sec-PSG}, the concept of projective symmetry group (PSG) is introduced. Here, the PSG on the star lattice is applied to the Sp($N$) mean field theory and is used to analyze possible $Z_2$ spin liquid phases. In Sec.~\ref{sec-phase}, various physical properties of two distinct $Z_2$ spin liquid phases are explained and the mean-field phase diagram including dimerized and magnetically ordered phases (for the nearest-neighbor Heisenberg model) is obtained. We discuss the implications of our results to theory and experiment in Sec.~\ref{sec-conclusion}. Details of the derivation of the PSG for the star lattice are given in Appendix \ref{sec-algebraic}.

\section{An Sp($N$) generalization of the Heisenberg model}
\label{sec-meanfield}

To investigate possible magnetically ordered and quantum paramagnetic states in the quantum antiferromagnetic
Heisenberg model, $H = \sum_{ij} J_{ij} \ \vec S_i \cdot \vec S_j$, it is useful to generalize the usual spin-SU(2) Heisenberg model to an Sp(N) model.\cite{sachdev-Kagome,coleman}

Let us start with the Schwinger boson representation of the spin operator ${\vec S}_i = b^{\dagger}_{i \alpha} {\vec \sigma}_{\alpha \beta} b_{i \beta}$, where $\alpha,\beta = \uparrow, \downarrow$, ${\vec \sigma}$ are Pauli matrices, $b_{i \alpha}$ are canonical boson operators and a sum over repeated $\alpha$ indices is assumed. Note that we need to impose the constraint $n_b = b^{\dagger}_{i \alpha} b_{i \alpha} = 2S$ to satisfy the spin commutation relations, where $S$ is the spin quantum number. A generalized model is obtained by introducing $N$ flavors of such bosons on each site. In order to keep the physical Hilbert space of spins, a constraint on the number of bosons given by $n_{b} = b^{\dagger m}_{i \alpha} b^{m}_{i\alpha} = 2S_\text{eff} = \kappa N$ where $m = 1,...,N$ must be imposed at each site. Note that $N=1$ corresponds to the physical limit Sp(1) $\equiv$ SU(2). The action of the corresponding Sp($N$) generalized model is then given by 
\begin{eqnarray}
{\cal S} = \int  d\tau \{{\bar b}_{i \alpha}^{m} \partial_{\tau} b_{i\alpha}^{m}
-\frac{J_{ij}}{2 N} {\bar A}_{ij} A_{ij} + \lambda_i (\-{b}_{i\alpha}^m b_{i\alpha}^m - n_{b}) \},
\label{genspn}
\end{eqnarray}
%\begin{eqnarray}
%\mathcal{S} &=& \int_0^\beta d\tau \{\-{b}_{i \alpha}^{m} \partial_{\tau} b_{i\alpha}^{m}
%-\frac{1}{2 N} \sum_{i,j} J_{ij} \-{A}_{ij} A_{ij} \\ \nonumber
%&+& \sum_i \lambda_i (\-{b}_{i\alpha}^m b_{i\alpha}^m - n_b) \}
%\label{genspn}
%\end{eqnarray}
where $A_{ij} = \epsilon_{\alpha \beta}~\delta_{m m'} b_{i\alpha}^{m} b_{j\beta}^{m'}$
($\epsilon_{\alpha\beta}~\delta_{m m'}$ is the Sp($N$) generalized antisymmetric tensor of SU($2$))
and the chemical potential $\lambda_i$ keeps the number of bosons fixed to
$n_b = \kappa N$ at every site. The mean-field action is then obtained by decoupling the quartic boson
interaction in $\mathcal{S}$ using the Hubbard-Stratonovich fields $Q_{ij} = -Q_{ji}$ directed
along the lattice links so that one obtains $Q_{ij} = \langle A_{ij} \rangle / N$ at the saddle point.
The mean field solution becomes exact in the large-$N$ limit where $N \rightarrow \infty$ is
taken while $\kappa = n_b/N$ is fixed.
We also introduce the parametrization
$b_{i\alpha}^{m} = \left( \begin{array}{c} \sqrt{N} x_{i\alpha}~~b_{i\alpha}^{\tilde{m}} \end{array} \right)^T$
where $\tilde{m}=2,...,N$ to allow for the possibility of long-range order that occurs when $x_{i\alpha} \not= 0$.
Consequently, after integrating over the bosons, 
we obtain the effective action at the large-$N$ saddle point (or the mean-field free energy) at zero temperature:
\begin{eqnarray}
{\cal S}_{\rm eff}/N &=& \sum_{i,j} \frac{J_{ij}}{2} (|Q_{ij}|^2
- Q_{ij} (\epsilon_{\alpha\beta} x^*_{i\alpha} x^*_{j\beta}) + c.c.) \nonumber \\
&+& \lambda \sum_i (|x_{i\alpha}|^2 - ({\kappa}+1))
+ \sum_\mu \omega_\mu(Q,\lambda) ,
\label{effaction}
\end{eqnarray}
where $\omega_{\mu} (Q,\lambda)$ are the eigenvalues of the mean-field Hamiltonian.
Note that
the chemical potential is now taken to be uniform since each site has the same number of nearest
neighbor links.
In general, magnetic ordering $x_{i \alpha} \not= 0$ occurs in the semiclassical limit
at larger $\kappa$
while quantum paramagnetic phases are obtained when $\kappa$ is small.
In this work, we will study possible phases of such a model as a function of
$\kappa$ and $J_d/J_t$ at zero temperature.

\section{Projective symmetric group analysis of $Z_2$ spin liquid phases on the star lattice}
\label{sec-PSG}

We are interested in the classification of Schwinger boson mean-field states, especially the spin liquid phases
that do not break any underlying microscopic symmetry. Such {\it symmetric} spin liquid phases can be
classified using a projective symmetry group analysis, which was previously used for the fermion\cite{wen1} and boson\cite{fawang} mean-field states for different lattices. For our purpose, the approach taken by Wang and Vishwanath\cite{fawang} would be the most relevant. This analysis allows us to identify all the physically realizable spin liquid phases, independent of particular microscopic Hamiltonians. In this section, we only consider the physical $N=1$ case of the Schwinger boson theory and note that distinct spin liquid phases may be realized as ground states in different models.

In the Schwinger boson theory, the effective action and all physical observables are invariant under the following local $U(1)$ transformation for the boson and mean-field ansatz $Q_{ij}$:
\beq
b_{i\alpha} & \rightarrow & e^{i\phi (i)} b_{i\alpha} , \cr
Q_{ij} & \rightarrow & e^{-i\phi(i)-i\phi(j)}Q_{ij} ,
\label{eq:Q}
\eeq
where $\phi (i)$ is an arbitrary real field defined on the underlying lattice site.
Therefore, two mean-field ansatze that are related by such a transformation correspond to
the same physical state after projection (onto the physical Hilbert space).
An important point is that symmetry transformations (such as space group, spin rotation,
and time reversal) may return a mean-field ansatz to a $U(1)$ transformed form and
in this case the transformed ansatz would correspond to the same physical state.
Thus when we consider the mean-field ansatz that preserves all the microscopic symmetries,
we need to include the $U(1)$ transformations.
The main idea of the projective symmetry group analysis is that a mean-field ansatz
preserves all the symmetries not only when the ansatz is invariant under the symmetry
transformation $X$, but also when it is invariant under the symmetry transformation $X$ followed
by a local $U(1)$ gauge transformation, $G_{X}$, {\it i.e.}
\beq
\left( G_X \cdot X \right) Q_{ij} = Q_{ij} .
\label{gx}
\eeq
Thus, for example, physically distinct symmetric spin liquid phases can be characterized by
different allowed sets of combined transformations, $\{ G_X \cdot X \}$.
%The set of all transformations that leave a mean-field ansatz invariant is called the projective
%symmetry group.

In addition, there also exist pure local gauge transformations that leave
the mean-field ansatz invariant. The set of such elements is called the invariant gauge group (IGG).
The IGG is a subgroup of the underlying $U(1)$ symmetry and is not a physical symmetry since it is
not related to any microscopic symmetry. On the other hand, the IGG becomes the emergent gauge symmetry in the deconfined phase that describes the relevant spin liquid phases.\cite{q_order}
Therefore, it is important to identify the IGG of a mean-field ansatz. The IGG and the set $\{ G_X \cdot X\}$ together form the PSG. This PSG then can be used to classify the physically distinct spin liquid phases that have the same microscopic symmetries.

It can be readily seen that the IGG of the mean-field ansatz $Q_{ij}$ on the star lattice
(or on any frustrated lattice) is $Z_2$. The two elements of the IGG are the identity operation
{\bf 1} and the IGG generator $-{\bf 1}$: $b_{i \alpha} \rightarrow -b_{i \alpha}$.
The spin liquid phases that are characterized by a $Z_2$ IGG are called $Z_2$ spin liquid states.
Here we would like to classify possible symmetric $Z_2$ spin liquid phases on the star lattice
using the PSG.

\subsection{Algebraic constraints on the PSG}
We would like to find all the constraints on the PSG that preserve microscopic symmetries such as
the space group, spin rotation, and time reversal. The Schwinger boson mean-field Hamiltonian
is explicitly spin-rotation invariant. 
Here we concentrate on the space group operations such as translations and point group
operations for the star lattice. 
The time reversal operation will be considered later. For each space group operation, the 
allowed gauge transformations in the PSG are strongly constrained by certain algebraic relations 
among symmetry group elements.
Thus we first need to derive all the algebraic relations (so-called algebraic PSGs) and
investigate the solutions which provide all the symmetric spin liquid phases.

In the case of the star lattice, the underlying Bravais lattice is a triangular lattice and
the space group contains two translations $T_1$ and $T_2$ defined by the basis vectors $\vec e_1$ and
$\vec e_2$ in Fig. \eqref{star}, one reflection $\sigma$ along the diagonal, and
the $60^\circ$ rotation $R$ about a lattice site.

The translation operation, $T_i$, shifts the lattice by one unit cell along $\vec e_i$,
\begin{subequations}
\label{translation}
\begin{align}
T_1:(r_1,r_2,\alpha_s)&\rightarrow(r_1+1,r_2,\alpha_s),\\
T_2:(r_1,r_2,\alpha_s)&\rightarrow(r_1,r_2+1,\alpha_s),
\end{align}
\end{subequations}
where ${\bf r} = (r_1, r_2, \alpha_s)$ represents the location of a lattice site.
Here $(r_1,r_2)$ with integers $r_1$ and $r_2$ denotes the coordinate of a unit cell
($\vec R =r_1 \vec e_1 + r_2 \vec e_2$) and $\alpha_s \in \{a,b,c,d,e,f\}$ labels the six sites
within each unit cell (see Fig.\eqref{star}).
Reflection $\sigma$, however, interchanges the sublattice indices,
\begin{subequations}
\label{reflection}
\begin{align}
\sigma:(r_1,r_2,a)&\rightarrow(r_2,r_1,e) ,\\
\sigma:(r_1,r_2,b)&\rightarrow(r_2,r_1,f) ,\\
\sigma:(r_1,r_2,c)&\rightarrow(r_2,r_1,d) ,\\
\sigma:(r_1,r_2,d)&\rightarrow(r_2,r_1,c) ,\\
\sigma:(r_1,r_2,e)&\rightarrow(r_2,r_1,a) ,\\
\sigma:(r_1,r_2,f)&\rightarrow(r_2,r_1,b) .
\end{align}
\end{subequations}
Rotation, $R$, also leaves the 6 sublattice indices interchanged,
\begin{subequations}
\label{rotation}
\begin{align}
R:(r_1,r_2,a)&\rightarrow(r_1-r_2,r_1,d) ,\\
R:(r_1,r_2,b)&\rightarrow(r_1-r_2,r_1,f) ,\\
R:(r_1,r_2,c)&\rightarrow(r_1-r_2,r_1,e) ,\\
R:(r_1,r_2,d)&\rightarrow(r_1-r_2-1,r_1,b) ,\\
R:(r_1,r_2,e)&\rightarrow(r_1-r_2-1,r_1,a) ,\\
R:(r_1,r_2,f)&\rightarrow(r_1-r_2-1,r_1,c) .
\end{align}
\end{subequations}

One can define the corresponding gauge transformation $G_{X}$ for each symmetry
operation $X = T_1, T_2, \sigma, R$:
\beq
G_{X}: b_{{\bf r} \alpha} \rightarrow e^{i \phi_X({\bf r})} b_{{\bf r} \alpha}.
\eeq
The PSG is then generated by combining the $Z_2$ IGG and the operations $G_X \cdot X$.
We follow Ref.\onlinecite{fawang} for the derivation of the algebraic relations between
the PSG elements that would impose strong constraints on possible spin liquid phases
and repeat some of the basic arguments here for completeness.

In order to see how the structure of the space group imposes the constraints on the PSG,
let us first consider the symmetry operation $T_1^{-1}T_2T_1 T_2^{-1}$ which is the identity
operation:
\beq
\label{tran-1}
T_1^{-1} T_2 T_1 T_2^{-1}: (r_1,r_2,\alpha_s) \rightarrow (r_1,r_2,\alpha_s),
\eeq
on every site. It means that the corresponding PSG operations should leave the mean-field ansatz unchanged, namely,
\beq
\left( G_{T_1} T_1 \right)^{-1} \left( G_{T_2} T_2 \right) \left( G_{T_1} T_1 \right) \left( G_{T_2} T_2 \right)^{-1}\in\text{IGG}.
\eeq
The PSG operation above can be rewritten as
$[T^{-1}_1(G_{T_1})^{-1} T_1] \cdot [T^{-1}_1 G_{T_2} T_1] \cdot [(T^{-1}_1T_2 G_{T_1} (T^{-1}_1 T_2)^{-1}] \cdot (G_{T_2})^{-1}$.
Since the gauge transformation $Y^{-1} G_X Y$ with a space group operation $Y$ acting on a site ${\bf r}$
would generate a phase $\phi_X(Y({\bf r}))$ in the boson field, the equation above leads to
the following constraint
\begin{eqnarray}
\label{eq-T2}
&-&\phi_{T_2}(\vec r) + \phi_{T_1}[T_2^{-1} T_1(\vec r)] \cr
&&+ \phi_{T_2}[T_1(\vec r)]  -\phi_{T_1}[T_1(\vec r)]  = p_1\pi ,
\end{eqnarray}
where $p_1=0,1$ comes from the fact that there are two elements, ${\bf 1}$ and $-{\bf 1}$ in
the IGG.

There are additional constraint equations from other independent space group operations. More specifically, together with Eq. \eqref{tran-1}, the following symmetry relations need to be
taken into account:
\begin{subequations}
\label{algebraic}
\begin{align}
T_2 T_1 &= T_1 T_2,\\
T_1 \sigma &= \sigma T_2,\\
\sigma^2 &=1,\\
T_1 R T_2&= R,\\
T_2 R &= R T_1 T_2,\\
\sigma R \sigma R&= I,\\
R^6 &=1.
\end{align}
\end{subequations}
It can be shown that all other relations can be derived from them.
In the Appendix ~\ref{sec-algebraic}, we solve all the algebraic constraints derived from these relations.
The general solution of the algebraic PSG for the star lattice is found as follows:
\begin{subequations}
\begin{align}
\phi_{T_1}(r_1,r_2,\alpha_s)&=0 ,\\
\phi_{T_2}(r_1,r_2,\alpha_s)&=p_1 \pi r_1,\\
\phi_{\sigma}(r_1,r_2,\alpha_s)&=p_1\pi r_1 r_2+\frac{p_2 \pi}{2} ,\\
\phi_R(r_1,r_2,\alpha_s) &= p_1 \pi r_1 r_2 + \frac{p_1 \pi}{2}r_2 (r_2 - 1)+ \frac{p_3\pi}{2} + p_4\pi
\delta_{\alpha_s,f}.
\end{align}
\end{subequations}
where $p_1,p_2,p_3,p_4\in\{0,1\}$. Here, $\delta_{\alpha_s,f}=1$ when $\alpha_s=f$ and zero otherwise. Thus there exist 16 possible symmetric spin liquid phases. Notice that, not surprisingly, the solutions for the translation and reflection are the same as those in the triangular lattice. The solution for the rotation, however, has a more complex structure.
%Notice that the gauge transformation $G_X$ associated with each symmetry transformation $X$ can be
%chosen such that they are independent of the sublattice index, $\alpha_s$ (see Appendix A).
The general solution, except for the rotation, looks similar to the one in the triangular and Kagome lattice cases where the underlying Bravais lattice is the same but the number of sites per unit cell is different. However, we will show later that once we consider mean-field ansatz with nonvanishing nearest-neighbor bond amplitudes, $Q_{ij}$, only two of these spin liquid phases survive and the properties of these states are different from the allowed states in the other cases.

To summarize this section, we solve the algebraic PSG constraint equations for the star lattice and find that $p_1,p_2,p_3,p_4 \in \{0, 1\}$ are required to classify all distinct symmetric $Z_2$ spin liquid states that preserve all space group symmetries. In the next section, we show that if the nearest-neighbor bond amplitudes $Q_{ij}$ are nonzero and time reversal invariance is required, there are additional constraints on these parameters. At a result, we will see that there exist only two symmetric $Z_2$ spin liquid states with distinct PSG or quantum order.

\subsection{$Z_2$ spin liquid states with nonvanishing nearest-neighbor bond amplitudes}

\begin{figure}[t]
\centering
\includegraphics[width=7.5cm]{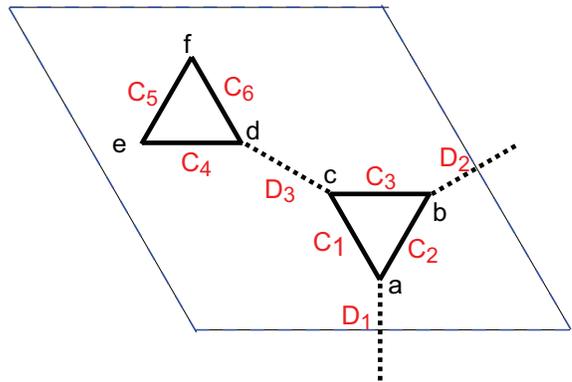}
\caption{(color online) In each unit cell, there are 9 different nearest-neighbor valence-bond amplitude, $Q_{ij}$, which can be classified into two groups,
$\{C_1, C_2, C_3, C_4, C_5, C_6 \}$ and $\{D_1, D_2, D_3\}$.}
\label{amplitude}
\end{figure}

In the star lattice, there are 9 different nearest-neighbor bond amplitudes
$Q_{ij}$ in the unit cell and we label them by
$\{C_1,\ldots,C_6\}$ and $\{D_1,D_2,D_3\}$ that correspond to the
triangular and bridge links, respectively (see Fig.\eqref{amplitude}).
If we assume that all of them are nonzero, there are more constraints 
on the PSG structure.

First, we consider what happens to the amplitude $D_3 (0,0) \equiv
Q_{(0,0,c) \rightarrow (0,0,d)}$ (other amplitudes are defined in a similar fashion) 
under reflection $\sigma$. From Eq.\eqref{reflection}, we infer
$D_3(0,0)\xrightarrow{\sigma} -D_3(0,0)$, then from the definition of
the PSG in Eq.\eqref{gx}, we get the constraint 
$\phi_\sigma(0,0,c)+\phi_\sigma(0,0,d)=\pi$ (mod $2\pi$).
This leads to $p_2=1$.

Another constraint can be obtained by comparing the $60^\circ$ rotation $R$, and
the reflection $\sigma$ on $C_1(0,0)$,
\beq (G_R \cdot R)C_1(0,0)=(G_\sigma \cdot  \sigma)C_1(0,0).
\eeq
Since $C_1(0,0)\xrightarrow{R} C_4(0,0)$ and $C_1(0,0)\xrightarrow{\sigma}
-C_4(0,0)$, it imposes the condition
\beq
\phi_\sigma^d+\phi_\sigma^e=\pi+\phi_R^d+\phi_R^e ,
\eeq
where $\phi_X^\alpha\equiv \phi_X(0,0,\alpha)$ for $X\in\{\sigma,R\}$.
This implies that $p_3=0$ in the PSG.

Finally, we consider the constraint by $180^\circ$ rotation $R^3$, and the translation $(T_1)^{-1}$ on the bridge link, $D_1(0,0)$,
\beq
(G_R \cdot R)^3 D_1(0,0) = (G_{T_1} T_1)^{-1} D_1(0,0) .
\eeq
Here, $D_1(0,0)\xrightarrow{R^3} -D_1(-1,0)$ and $D_1(0,0)\xrightarrow{T_1^{-1}} D_1(-1,0)$, leading to the constraint,
\beq
\phi_R^c(0,0)&+&\phi_R^d(0,0)+\phi_R^e(0,0)+\phi_R^b(-1,0)\nonumber\\
&+&\phi_R^a(-1,0)+\phi_R^f(-1,-1)=\pi ,
\eeq
which implies $p_4=1$ in the PSG.

\begin{figure}[t]
\centering
\includegraphics[width=8cm]{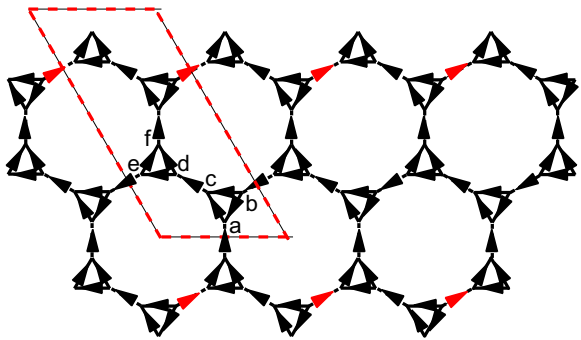}
\caption{(color online) The arrow representation of the mean-field ansatz for the $p_1=1$ state. 
The area enclosed by the dashed lines is the corresponding unit cell with 12 sites.
Note that the directed link $Q_{be}$ is staggered along the $(0,1)$ direction.}
\label{p1=1}
\end{figure}

\begin{figure}[t]
\centering
\includegraphics[width=8cm]{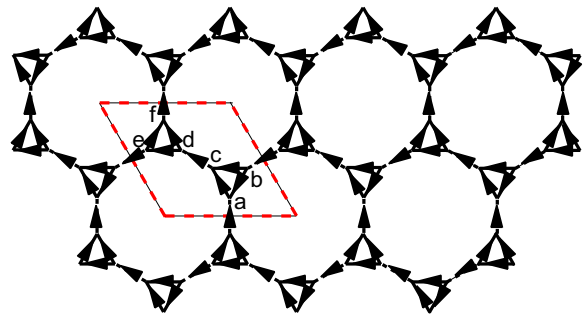}
\caption{(color online) The arrow representation of the mean-field ansatz for the $p_1=0$ state. 
The arrow from the site $i$ to site $j$ means $Q_{ij} > 0$. The area enclosed by the dashed lines is 
the corresponding unit cell with 6 sites.}
\label{p1=0}
\end{figure}

Thus, by assuming nonvanishing nearest-neighbor amplitudes, the
parameters which characterize the PSG structure
$\{p_1,p_2,p_3,p_4\}$ become $\{p_1,1,0,1\}$. There are only two distinct 
symmetric $Z_2$ spin liquids corresponding to $p_1=0,1$. If
the time reversal symmetry is preserved, all the amplitudes 
$Q_{ij}$ can be taken to be real. Moreover, $Q_{ij} = - Q_{ji}$ that follows from the self-consistent equation. Hence the mean field ansatz ${Q_{ij}}$ can be depicted by an arrow representation in which the arrow denotes the direction where $Q_{ij}$ is taken to be positive. The arrow representations for the two distinct spin liquid phases are shown in Fig. \eqref{p1=1} and \eqref{p1=0} respectively. The 
$p_1=1$ state can be described by a unit cell with 12 sites while $p_1=0$ state has a unit cell with 6 sites. Both of them are characterized by two kinds of nearest-neighbor bond amplitudes 
$Q_d$ and $Q_t$, which refer to the {\it absolute} values, $|Q_{ij}|$, of the amplitudes on the bridge and triangular links, respectively.

These two states can also be distinguished by the ``flux" enclosed in a length-12 polygon\cite{oleg}, which is defined as the phase $\Theta$ of the gauge-invariant product of the nearest-neighbor amplitudes along the 12-length loop: 
\beq Q_{i_1i_2} (-Q_{i_2i_3}) \cdots (-Q_{i_{12}i_1}) = Q_d^6 Q_t^6 e^{i\Theta} ,
\eeq
where $\{i_1, \ldots, i_{12} \}$ label the 12 sites along a lenght-12 loop as shown in Fig. \eqref{star}. The ``flux" $\Theta=0$ for $p_1=1$ state (zero-flux state) while $\Theta=\pi$ for $p_1=0$ state ($\pi$-flux state). Hence, the two states are clearly not gauge-equivalent and can be identified by the ``flux". The physical properties of these spin liquid states and how they arise in the large-$N$ Sp($N$) mean-field theory will be discussed in the next section.

\section{Large-$N$ Sp($N$) Mean-field Phase Diagram}
\label{sec-phase}
In this section, we analyze the large-$N$ mean-field theory of the Sp($N$)-generalized Heisenberg model with the nearest-neighbor exchange interactions. In particular, we investigate the phase diagram as a function of $J_d/J_t$ and $\kappa = 2 S_{\rm eff}$. In the previous section, we demonstrate that there are only two possible symmetric spin liquid phases, as shown in Fig.\eqref{p1=1} and Fig.\eqref{p1=0}, 
when the nearest-neighbor bond amplitudes are finite and they correspond to $p_1=1,0$ in the PSG description respectively. The strength of the nearest-neighbor bond amplitudes, $Q_d$ and $Q_t$, and the spinon condensate density $x_{i\alpha}$ can be determined by minimizing the effective action, Eq. \eqref{effaction}.

\begin{figure}[t]
\centering
\includegraphics[width=7cm,angle=270]{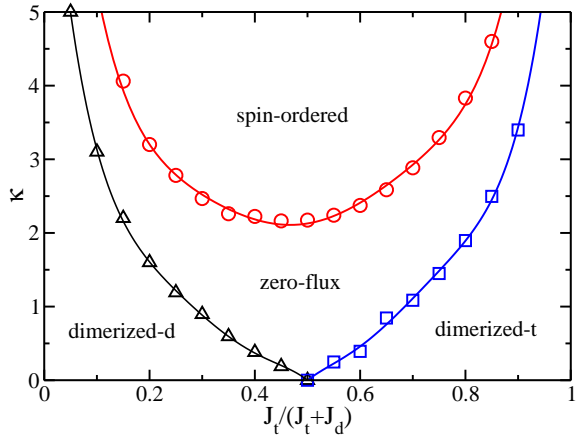}
\caption{(color online) Large-$N$ mean-field phase diagram for the Sp($N$) generalized nearest-neighbor Heisenberg model. Notice that only the zero-flux state occurs as a stable spin liquid state in the phase diagram.}
\label{phase-p1=1}
\end{figure}

\begin{figure}[t]
\centering
\includegraphics[width=7cm,angle=270]{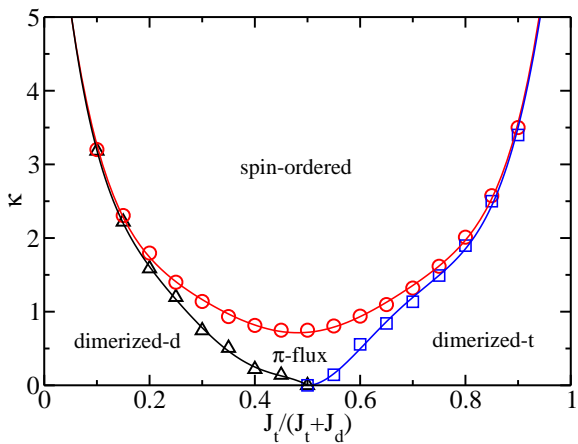}
\caption{(color online) Large-$N$ mean-field phase diagram with the zero-flux state being artificially suppressed (as if an additional interaction punishes the zero-flux state). The $\pi$-flux state is then the spin liquid state competing with the dimerized states.}
\label{phase-p1=0}
\end{figure}

In the Sp($N$)-generalized Heisenberg model, it has been known that the spin liquid state with the smallest ``flux" has the lowest energy. Thus, not surprisingly, we find that the zero-flux state ($p_1=1$) is always lower in energy in the relevant part of the phase diagram. On the other hand, it is also known that a ring-exchange or the next-nearest-neighbor spin interactions can lower the energy of 
a spin liquid state with a larger flux \cite{fawang}. Hence, it is useful to analyze the phase diagram of the Heisenberg model with respect to both of the two spin liquid states. The mean-field phase diagram for the nearest-neighbor model is shown in Fig.\eqref{phase-p1=1}, where the $\pi$-flux state never appears as the true ground state. Anticipating that other types of interactions can favor the $\pi$-flux state, we also compute the mean-field phase diagram by artificially suppressing the zero-flux state (as if an appropriate additional interaction may punish the zero-flux state). The resulting phase diagram is shown in Fig.\eqref{phase-p1=0}. Notice that the magnetically ordered phases in the large-$\kappa$ limit in Fig.\eqref{phase-p1=1} and Fig.\eqref{phase-p1=0} are descendants of the zero-flux and $\pi$-flux phases in the sense that the condensation of the spinons in each spin liquid state leads to
these magnetically ordered phases. On the other hand, the ground states in the $\kappa \ll 1$ limit are typically dimerized or valence bond crystal phases. Physical properties of all the phases present in the phase diagram and the interplay between them are described as follows. 

%\begin{figure}
%\centering
%\includegraphics[width=8cm]{exhop.eps}
%\includegraphics[width=4cm]{exhop2.eps}
%\caption{(Left) VBS$_d$. (Right) VBS$_t$}
%\label{VBS}
%\end{figure}

\subsection{zero-flux spin liquid state and the related magnetically ordered phase}
The zero-flux spin liquid state corresponds to $p_1=1$ in the PSG description and the mean-field ansatz is shown 
in Fig.\eqref{p1=1}. There are 12 sites per unit cell in the mean-field ansatz. It has zero flux in the 12-sided and $\pi$-flux in the 14-sided polygon (see Fig.\eqref{star}). Hence, it has the lowest energy in the pure Heisenberg model according to the flux expulsion argument by Tchernyshyov et al.\cite{oleg} in the small $\kappa$ limit. It is an analogous state of the [0Hex,$\pi$Rhom] spin liquid state identified in the Kagome lattice.\cite{fawang}

\begin{figure}[t]
\centering
\includegraphics[width=4cm]{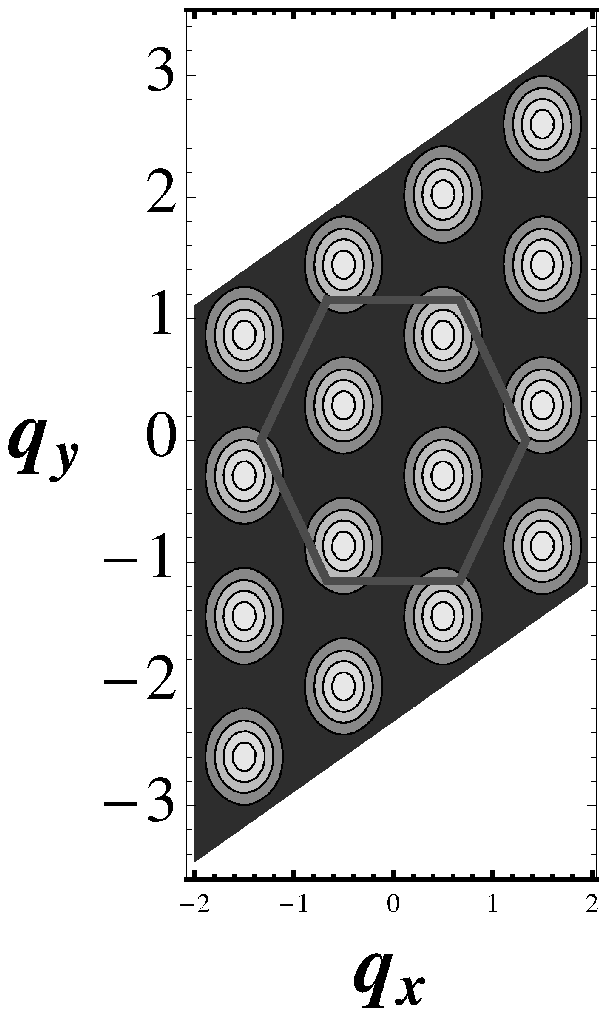}
\includegraphics[width=4cm]{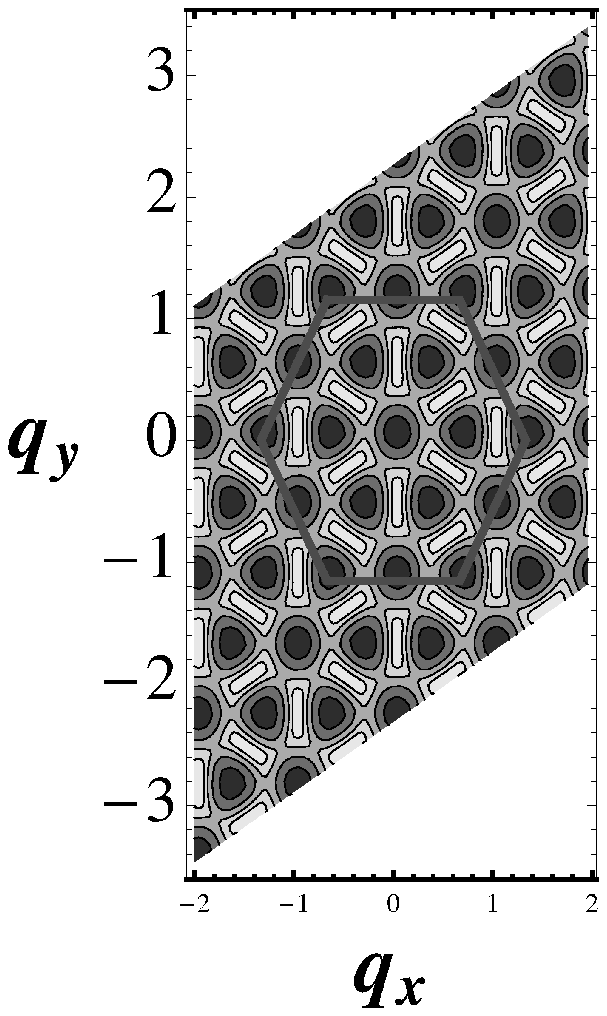}
\caption{Contour plots of the single-spinon (left) spectrum and the lower edge of the two-spinon/spinon-antispinon (right) spectrum of 
the zero-flux state ($p_1=1$). Darker area means lower energy. The hexagon represents the Brillouin zone.}
\label{spinon-1}
\end{figure}

The spinon spectrum can be computed using the Sp($N$) theory described in section ~\ref{sec-meanfield}. However, the single-spinon spectrum is not gauge-invariant and the gauge-invariant two-spinon (spinon-antispinon) spectrum is physically more relevant. Here, we present the lower-edge of the two-spinon spectrum, which is given by
\beq
E^{(2)} (\vec q) = \min_{\vec p} \{\epsilon_{\vec q-\vec p} + \epsilon_{\vec p}\} ,
\eeq
where $\epsilon_{\vec p}$ is the single-spinon spectrum. The single-spinon spectrum and the lower edge of the two-spinon spectrum are shown in Fig.\eqref{spinon-1} which are similar to that of [$0$Hex,$\pi$Rhom] spin liquid phase obtained in Kagome lattice.\cite{fawang} The minima of two spinon spectrum are given by $\vec q=\pm(\pi/3+n\pi,\pi/3+m\pi)$ and $(n\pi,m\pi)$ with integer $n,m$. As $\kappa$ increases, the minimum of the spinon spectrum decreases and the spectrum becomes gapless at $\kappa=\kappa_c$, where $\kappa_c=\kappa_c (J_d/J_t)$ varies depending on the value of $J_d/J_t$. Possible magnetically ordered phases arising when $\kappa > \kappa_c = \kappa_c (J_d/J_t)$ are characterized by the ordering wavevectors $\vec q=\pm(\pi/3+n\pi,\pi/3+m\pi)$ and $(n\pi,m\pi)$ with integers $n,m$. 

%which include the $\sqrt 3\times \sqrt 3$ state and other nonplanar {\bf (Hmm ... WE SHOULD CHECK THIS)} states.\cite{sachdev-Kagome}

\subsection{$\pi$-flux spin liquid state and the related magnetically ordered phase}
\begin{figure}[t]
\centering
\includegraphics[width=4cm]{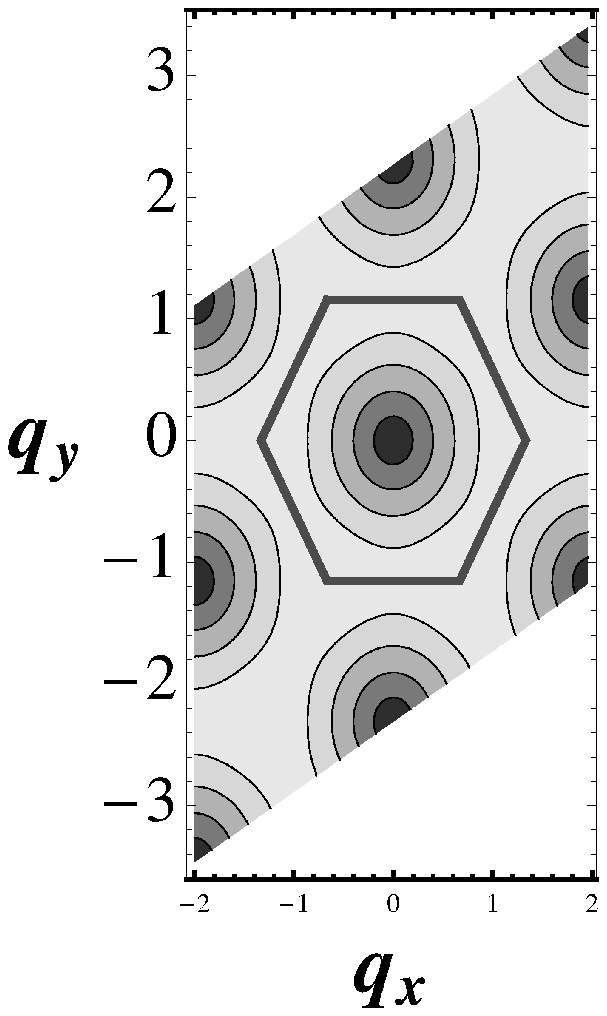}
\includegraphics[width=4cm]{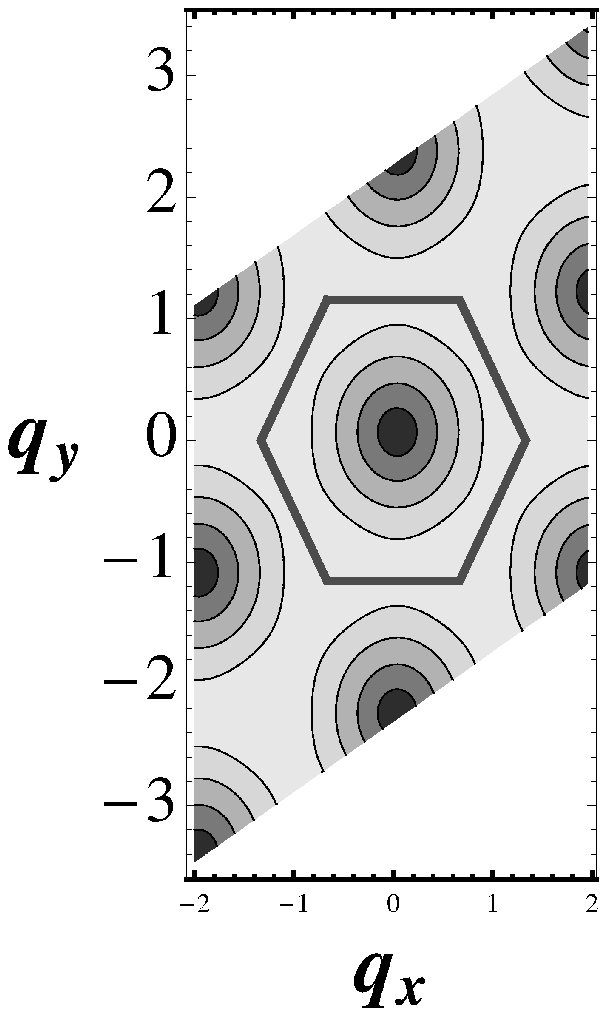}
\caption{Contour plots of the single-spinon (left) spectrum and the lower edge of the two-spinon/spinon-antispinon (right) spectrum of 
the $\pi$-flux state ($p_1=0$). Darker area means lower energy. The hexagon represents the Brillouin zone.}
\label{spinon-0}
\end{figure}

The $\pi$-flux spin liquid state is characterized by $p_1=0$ in the PSG description and the mean-field ansatz (in the arrow representation) is depicted in Fig.\eqref{p1=0}. The ansatz is described by a 6-site unit cell. It has $\pi$ flux in the 12-sided and zero flux in the 14-sided polygon as shown in Fig. \eqref{star}. Both single- and two- spinon spectrum are shown in Fig.\eqref{spinon-0}. The two-spinon spectrum has the global minimum at the center of the Brillouin zone $\vec q=(0,0)$. It is an analogous state of the $Q_1=Q_2$ state identified in the Kagome lattice.\cite{sachdev-Kagome,fawang} The condensation of the spinons leads to the $q=0$ magnetically ordered ground state which is translationally invariant. Since the two-spinon spectrum of the $\pi$-flux state is quite different from that of the zero-flux state, the two states can be distinguished by neutron scattering experiment that measures spin-1 excitations.

\subsection{Dimerized-d state}
In the regime $J_t<J_d$, the ground state is a dimerized state for sufficiently small $\kappa$,
where all the triangular bond amplitudes vanish ($Q_t=0$) and only the amplitude on the bridge
links, $Q_d$, is finite. We call this state as the dimerized-d state. 
Notice that this state does not break any translational symmetry.
This is an isolated-dimer state and there is a gap $\sim J_d$ for the spin-1 excitations. 
The presence of this ground state in the small $\kappa$ limit can be proven using the small $\kappa$
expansion\cite{oleg} of the effective action $S_{\rm eff}$ in Eq.\eqref{effaction} for paramagnetic solutions ($x_{i \alpha}=0$). 
Such a perturbative expansion of $S_{\rm eff}$ in $\kappa$ leads to 
\beq \label{eq:S}
\frac{S_{\text{eff}}}{NN_s}=-\frac{P_1}{R}\kappa-\frac{P_2}{2 R
P_1}\kappa^2+O(\kappa^3) ,
\eeq
where $N_s$ is the number of lattice sites, 
$R \equiv -(J_d Q_d^2 + 2 J_t Q_t^2)/2$, and $P_n$ is the ``flux operator" defined on the loop of length $2n$,
\beq
P_n \equiv \frac{1}{N_s} \sum_{\text{loop}} (\frac{J_{12}}{2} 
Q_{12} ) (- \frac{J_{23}}{2} Q_{23}^* ) \cdots ( - \frac{J_{2n,1}}{2} 
Q_{2n,1}^* ) .
\eeq
In particular, $P_1=-(J_d^2 Q_d^2 + 2 J_t^2 Q_t^2)/4$ and $P_2=(8J_d^2 
Q_d^2 J_t^2 Q_t^2 + 6 J_t^4 Q_t^4 + J_d^4 Q_d^4)/16$.

When $J_t<J_d$, we find that $Q_t=0$ can minimize the effective action Eq.\eqref{effaction} for 
$\kappa < \kappa^d_c$ and the critical $\kappa^d_c$ is 
\beq
\kappa^d_c = 2\frac{1-2u}{4 u-1} ,
\eeq
where $u \equiv J_t/(J_t+J_d) \leq 1/2$. This result is asymptotically correct near $u=1/2$
where $\kappa^d_c = 0$. When $\kappa > \kappa^d_c$, the spin liquid phases
become more stable as far as $\kappa$ is not too large.

\subsection{Dimerized-t state}
When $J_t > J_d$, there is another dimerized mean-field state for sufficiently small $\kappa$. 
In this state, all the amplitudes on the bridge links are zero ($Q_d=0$) while the amplitudes on
the triangular links are finite. The mean-field dimerized-t state, therefore, does not break any
translational symmetry. Again, the presence of this state can be seen from the small $\kappa$
expansion of the effective action. That is, $Q_d=0$ is the solution for the minimum effective action
as far as $\kappa < \kappa^t_c$, where $\kappa^t_c$ is
\beq
\kappa^t_c = 4\frac{2 u-1}{5-8u} .
\eeq
When $\kappa > \kappa^t_c$, the spin liquid phases become more stable for not-too-large $\kappa$.

\subsection{Further discussions on the phase diagram}
Notice that, at the isotropic point, $J_t=J_d$, a spin liquid phase becomes the ground state 
even in the small $\kappa$ limit, where the amplitudes on both the bridge and triangular links
are nonvanishing and identical (see Fig.\eqref{phase-p1=0} and Fig.\eqref{phase-p1=1}). 
However, it turns out that the amplitudes on the bridge links become stronger than the ones on the triangular links as $\kappa$ increases. This indicates a tendency to form local singlets on the bridge links, which may be consistent with the results of the exact diagonalization study by Richter et al.\cite{richter} for the spin-1/2 isotropic model. More precise determination of the ground state at the isotropic point, therefore, requires the analysis of $1/N$ fluctuations about the large-$N$ mean-field state.

The spin-1/2 anisotropic model with $J_d \not= J_t$ was previously studied by exact diagonalization restricted to the dimer Hilbert space.\cite{misguich} It was found that the dimerized-d state is the stable ground state for $J_t < 1.3 J_d$. On the other hand, for the opposite limit $J_t > 1.3 J_d$, it was suggested that the ground state may be a valence bond crystal made of a lattice of 18-sided plaquette-valence-bond structure, which breaks the translational symmetry and is three-fold degenerate.\cite{misguich} Our mean-field theory cannot capture possible presence of this state since such a state would arise via fluctuations beyond the large-$N$ limit.\cite{marston,VBS} Thus the incorporation of relevant quantum fluctuations or another method is necessary to  pin down the ultimate fate of the mean-field dimerized-t state.

As discussed in the previous sections, the mean-field transition from the spin liquid phases to magnetically ordered phases is continuous since it is described by the condensation of bosonic spinons. When $J_d$ and $J_t$ are not very different from each other, there is no direct transition from the dimerized state to magnetically ordered phases. On the other hand, in the extreme anisotropic limits, $J_d \gg J_t$ or $J_d \ll J_t$, there is a possibility in the $\pi$-flux phase diagram that there is a direct transition from a dimerized state to a magnetically ordered state - the energies of all the states become very close near the phase boundary so that our mean-field calculation could not determine whether there is a direct transition or one still has to go through a spin liquid phase in the extreme anisotropic cases. If a direct transition is possible, such a transition does not have to be always first order because the dimerized-d state, for example, does not break any spatial symmetry. The transition from the spin liquid phases to dimerized states (with isolated dimers) is continuous and is described by the confinement-deconfinement transition of spinons in a $Z_2$ gauge theory \cite{shastry_lattice}.

\section{Discussion}
\label{sec-conclusion}
In the large-$N$ mean-field phase diagram of the star-lattice Heisenberg model, it is found that the two possible $Z_2$ spin liquid phases can exist even for $\kappa > 1$ (this corresponds to $S>1/2$ in the physical $N=1$ limit) in some parts of the phase diagram. This is highly unusual given that most of the previous studies on other lattice models obtain $\kappa_c < 1$. While the phase boundaries in the large-$N$ mean-field theory may change as $N$ gets smaller, this is certainly an encouraging sign. Notice, for example, that, when $\kappa=3$ ($S_{\rm eff}=3/2$), as $J_t/(J_d+J_t)$ changes from zero to one, one encounters the dimerized-d state, zero-flux spin liquid, magnetically ordered state, zero-flux state and finally the dimerized-t state in the mean-field phase diagram (see Fig.\eqref{phase-p1=0}). 
%When only the $\pi$-flux state and related magnetically ordered state are considered, one can only get the dimerized-d state, $\vec q = 0$ magnetically ordered state, and then dimerized-t state in sequence. 

More generally, the zero-flux phase is stable up to a relatively large $\kappa = 2S_{\rm eff}$: $\kappa_c \sim 2$ at the uniform point ($J_d=J_t$). It is close to the $\kappa_c$ obtained in [$0$Hex,$\pi$Rhom] spin liquid phase in Kagome lattice, which has the similar two-spinon spectrum.\cite{fawang} The $\kappa_c$ gets even larger in the anisotropic limit, $J_t\gg J_d$ or $J_t\ll J_d$. The largest $\kappa_c$ we obtain is $\kappa_c \sim 5$ in a very anisotropic limit, $J_d/J_t\sim 9$. On the other hand, in such an anisotropic limit, the region in the phase diagram where the spin liquid state is stable becomes smaller. This suggests that moderately anisotropic exchange interactions may favor the realization of spin liquid phases.

It is also worthwhile to notice that the $\pi$-flux state has a relatively small $\kappa_c$, in contrast to a similar study of the Kagome lattice model\cite{fawang}. As emphasized in Ref.\onlinecite{fawang}, the spin liquid phases with finite flux may be stabilized by a ring exchange term that arise near a metal-insulator transition where charge fluctuations become important.\cite{motrunich} Thus the star-lattice antiferromagnetic insulator at the verge of becoming a metal may be a good candidate for the realization of the $\pi$-flux spin liquid state.

Finally, as far as we know, the only known realization of the star-lattice antiferromagnet is the polymeric Iron(III) Acetate [Fe$_3$($\mu_3$-O)($\mu$-OAc)$_6$(H$_2$O)$_3$][Fe$_3$($\mu_3$-O)($\mu$-OAc)$_{7.5}$]$_2\cdot$7H$_2$O.\cite{zheng} The spin of the magnetic Fe$^{\rm III}$ ion is $S=5/2$ and the Curie-Weiss temperature determined from the high temperature susceptibility is $\Theta_{\rm CW} = - 581$K. This material undergoes a magnetic transition at $T_N \sim 4.5$K, leading to a large frustration parameter, $f = |\Theta_{\rm CW}|/T_N = 129$. The magnetic ordering patterns predicted in the large $\kappa$ limit of the large-$N$ mean-field theory may directly be relevant to the low temperature phase of this system. Once the magnetic ordering pattern is determined by neutron scattering experiment or other means, one may be able to determine whether the material is close to the zero-flux or $\pi$-flux spin liquid phases because they are related to different magnetically ordered phases.\cite{isakov} The large frustration parameter observed in this material and the large $\kappa_c$ from our mean-field theory point to the possibility that a spin-1/2 or even a spin-1 analog of this material may support one of the spin liquid phases discussed in this work.

\acknowledgments
This work was supported by the NSERC of Canada, the Canadian Institute for Advanced Research,
and the Canada Research Chair Program.

\appendix
\section{Algebraic PSG for the star lattice}
\label{sec-algebraic}
Here we generalize the method developed in the Ref.\onlinecite{fawang} to derive the allowed PSGs for the 
star lattice. The strategy is to find all the constraints on the PSGs and use them to identify the general solution.
We first consider how PSG transforms under an arbitrary U(1) gauge transformation
$G\equiv e^{i\phi_G}$ on the ansatz, $Q_{ij}\rightarrow G Q_{ij}$. 
The transformed ansatz should now be invariant under $GG_X XG^{-1}=\left(G G_X
XG^{-1}X^{-1}\right)X$. Thus $G_X$ can be replaced by $G G_X XG^{-1}X^{-1}$.
This means that the phase transforms as
\beq
\phi_X(\vec r)\rightarrow \phi_G(\vec r) +\phi_X(\vec
r)-\phi_G\left(X^{-1}(\vec r)\right ) .
\eeq
Here $\vec r=(r_1,r_2,\alpha_s)$ with integers $r_1$ and $r_2$ which
label the location of the unit cell, $\vec R=r_1 \vec e_1+ r_2 \vec e_2$, and
$\alpha_s \in\{a,b,c,d,e,f\}$ label the six sites in a unit cell.

To simplify the expressions of the PSG, one can choose $\phi_{T_1}(r_1,r_2,\alpha_s)=0$ 
and $\phi_{T_2}(0,r_2,\alpha_s)=0$ (independent of the sublattice index $\alpha_s$), by using 
a gauge degree of freedom or the gauge transformation $G_0$ via
\beq \phi_{G_0}(r_1,r_2)=-\sum_{i=-\infty}^{r_1}
\phi^0_{T_1}(i,r_2) - \sum_{j=-\infty}^{r_2}\phi^0_{T_2}(0,j)
\eeq
on all sublattices $\alpha_s$. Here $\phi^0_{T_1}$ and $\phi^0_{T_2}$
correspond to the phases for an arbitrary initial choice for 
$G_{T_1}$ and $G_{T_2}$.
Notice that the gauge
transformation $G_0$ is well-defined only on the lattice with open
boundary condition. Extra care is necessary for periodic boundary
condition. We assume open boundary condition throughout 
the analysis for simplicity.

Now we would like to find the PSGs which satisfy all the algebraic 
constraints in Eq.\eqref{algebraic}. First,
we consider the constraint arising from the symmetry relation,
$T_1T_2=T_2T_1$, in Eq.\eqref{eq-T2}, 
\beq 
\Delta_1 \phi_{T_2}(\vec r)=p_1\pi ,
\eeq
where we introduce two forward difference
operators $\Delta_1$ and $\Delta_2$, defined as $\Delta_1
f(r_1,r_2)\equiv f(r_1+1,r_2)-f(r_1,r_2)$ and $\Delta_2 f(r_1,r_2) \equiv
f(r_1,r_2+1)-f(r_1,r_2)$. Here $p_1=0,1$ is a site-independent integer
corresponding to the two elements in IGG. The solution for
$\phi_{T_2}$ then is given by
\beq \label{t2}
\phi_{T_2}(r_1,r_2,\alpha_s)=p_1\pi r_1,
\eeq
which is independent of $\alpha_s$.

Next, we consider the relation, $\sigma T_2 = T_1\sigma$ and $\sigma
T_1 = T_2\sigma$. The constraints arising from these relations are
\begin{subequations}
\begin{align}
\Delta_1 \phi_\sigma(r_1,r_2,\alpha_s)&= p_2' \pi + p_1\pi r_2 ,\\
\Delta_2 \phi_\sigma(r_1,r_2,\alpha_s)&= p_3' \pi + p_1\pi r_1,
\end{align}
\end{subequations}
after substituting Eq. \eqref{t2} for $\phi_{T_2}(\vec r)$ and
$p_2',p_3'=0,1$. The solution to these equations is
\beq
\label{sigma1}
\phi_\sigma(r_1,r_2,\alpha_s)=\phi_\sigma^{\alpha_s}+p_2'\pi r_1 + p_3'
\pi r_2 + p_1 \pi r_1 r_2 ,
\eeq
where
$\phi_\sigma^{\alpha_s}\equiv\phi_\sigma(0,0,\alpha_s)$.
Here $p_2',p_3'$ and $\phi_\sigma^{\alpha_s}$ can further be determined by 
additional symmetry relations.

Notice that, from $\sigma\sigma=I$, we have
\begin{subequations}
\begin{align}
\phi_\sigma(r_1,r_2,a)+\phi_\sigma(r_2,r_1,e)=p_2\pi ,\\
\phi_\sigma(r_1,r_2,b)+\phi_\sigma(r_2,r_1,f)=p_2\pi ,\\
\phi_\sigma(r_1,r_2,c)+\phi_\sigma(r_2,r_1,d)=p_2\pi .
\end{align}
\end{subequations}
Again $p_2=0,1$ correspond to the two elements of the IGG, which is
sublattice-independent. Using Eq.\eqref{sigma1}, we get the following
constraint equation.
\beq
\phi_\sigma^a+\phi_\sigma^e=p_2\pi+(r_1+r_2)(p_2'+p_3')\pi ,
\eeq
and hence $p_2'=p_3'$ (modulo 2) because the left-hand-side is independent of $r_1$,
$r_2$. To determine $p_2'$, consider the gauge transformation $G_1$,
\beq
G_1: \phi_{G_1}(r_1,r_2,\alpha_s) = \pi r_1.
\eeq
One can show that the gauge transformation $G_1$ does not modify
$G_{T_1}$ and $G_{T_2}$, but $G_{\sigma}$ changes as follows:
\beq
\phi_\sigma(\vec r)\rightarrow \phi_\sigma(\vec r)=
\phi_\sigma^{\alpha_s}+(p_2'+1)\pi(r_1+r_2)+p_1\pi r_1 r_2.\nonumber
\eeq
Therefore we can always assume $p_2'=p_3'=0$ (modulo 2) and this leads to
\beq
\phi_\sigma(r_1,r_2,\alpha_s)=\phi_\sigma^{\alpha_s}+p_1 \pi r_1 r_2 .
\eeq
To determine $\phi_\sigma^{\alpha_s}$, we consider the following
gauge transformation,
\beq
G_2: \left\{ \begin{array}{cccc}
\phi_2(r_1,r_2,a)&=&\phi_0,&\\
\phi_2(r_1,r_2,e) &=&-\phi_0,&\\
\phi_2(r_1,r_2,\alpha_s) &=& 0 &\alpha_s \notin \{a,e\}.
\end{array}\right.
\eeq
where $\phi_0$ is an arbitrary constant. Again, this transformation does not change $G_{T_1}$ and
$G_{T_2}$, but modifies $G_\sigma(a)$ ($G_\sigma$ acting on the sublattice site $a$)
and $G_\sigma(e)$ as follows:
\beq
\phi_\sigma^a&\rightarrow& \phi_\sigma^a+2\phi_0\q,\\
\phi_\sigma^e&\rightarrow& \phi_\sigma^e-2\phi_0\q,\\
\phi_\sigma^\alpha&\rightarrow& \phi_\sigma^{\alpha_s}\q\alpha_s \notin
\{a,e\}.
\eeq
By choosing $\phi_0 =\frac{1}{4}(\phi_\sigma^e-\phi_\sigma^a)$, we can make the phases
$\phi_\sigma^a$ and $\phi_\sigma^e$ to be equal, {\it i.e.} 
$\phi_\sigma^a=\phi_\sigma^e=p_2\pi/2$. 
Similar gauge transformations, $G'_2$ and $G''_2$, can be used for 
$\{b,f\}$ and $\{c,d\}$ pairs such that all the phases, $\phi_\sigma^{\alpha_s}$, are chosen to be 
$p_2 \pi/2$.
The resulting PSG for the reflection, $G_\sigma$, is then given by
\beq \label{sigma} 
\phi_\sigma(r_1,r_2,\alpha_s)=p_1 \pi r_1 r_2 + \frac{p_2 \pi}{2} ,
\eeq
which is independent of the sublattice index, $\alpha_s$.

Now let us consider algebraic constraints arising from $T_1 R T_2 = R$ and $R T_1 T_2 = T_2 R$:
\begin{subequations}
\begin{align}
\Delta_1 \phi_R (r_1,r_2,\alpha_s) &= p_1\pi r_2  + p_4'\pi , \\
\Delta_2 \phi_R (r_1,r_2,\alpha_s) &= p_1\pi (r_1-r_2-1) + p_4 \pi ,
\end{align}
\end{subequations}
using the solution of $\phi_{T_1}$ and $\phi_{T_2}$. Here,
$p_4',p_4 =0,1$ can be fixed by using additional algebraic constraints 
and the gauge degrees of freedom. The general solution of the difference equations for $\phi_R(\vec r)$ is found as
\beq \label{rotation1}
\phi_R(r_1,r_2,\alpha_s) &=& p_1 \pi r_1 r_2 + p_4' \pi r_1 + p_4 \pi
r_2 \nonumber \\
&+& \frac{p_1 \pi}{2}r_2 (r_2 - 1) +\phi_R^{\alpha_s} ,
\eeq
where $\phi_R^{\alpha_s} \equiv \phi_R(0,0,\alpha_s)$. To determine $p_4'$, we
consider the relation $R\sigma R \sigma = I$ with Eq.\eqref{sigma} and Eq.\eqref{rotation1},
which results in, for example,
\beq
2\phi_R^a+p_4'\pi(r_2-1)=p_3\pi
\eeq
for $p_3=0,1$ and hence it implies $p_4' = 0$ (modulo 2). Similarly, we can obtain the following set of coupled equations,
\begin{subequations}
\begin{align}
2 \phi_R^a &=p_3 \pi ,\\
\phi_R^b + \phi_R^c &=p_3 \pi ,\\
\phi_R^d + \phi_R^e &=p_3 \pi ,\\
2 \phi_R^f &=p_3 \pi .
\end{align}
\end{subequations}
Unlike the case of the Kagome lattice, there is no further constraint
imposed by the relation $\sigma R\sigma R = I$. To fix the gauge degree of freedom for
$\phi_R^{\alpha_s}$, we consider another gauge transformation,
\beq
G_4 = \left\{ \begin{array}{cccc}
\phi_4(r_1,r_2,a)&=&\phi_1,&\\
\phi_4(r_1,r_2,e)&=&\phi_1,&\\
\phi_4(r_1,r_2,b)&=&\phi_2,&\\
\phi_4(r_1,r_2,f) &=&\phi_2,&\\
\phi_4(r_1,r_2,\alpha_s) &=& 0 &\text{otherwise}.
\end{array}\right.
\eeq
This gauge transformation does not modify $G_{T_1}$, $G_{T_2}$ and $G_\sigma$, but changes $\phi_R$:
\beq
\phi_R^b&\rightarrow& \phi_R^b+ \phi_2 ,\\
\phi_R^c&\rightarrow& \phi_R^c- \phi_2 ,\\
\phi_R^d&\rightarrow& \phi_R^d- \phi_1 ,\\
\phi_R^e&\rightarrow& \phi_R^e+ \phi_1 ,\\
\phi_R^{\alpha_s}&\rightarrow& \phi_R^{\alpha_s} \q \text{otherwise}.
\eeq
One can show that, by suitable choices of $\phi_1$ and $\phi_2$, all $\phi_R^{\alpha_s}$ can be made to be identical and equal to $p_3 \pi/2$. To simplify the terms that involve $p_4$, we consider another gauge transformation,
\beq
G_5 = \left\{ \begin{array}{cccc}
\phi_5(r_1,r_2,a)&=&\pi(r_1+r_2),&\\
\phi_5(r_1,r_2,e) &=&\pi(r_1+r_2),&\\
\phi_5(r_1,r_2,\alpha_s) &=&\pi(r_1+r_2+1)&\alpha_s \notin \{a,e\},
\end{array}\right.
\eeq
which does not modify $G_{T_1}$, $G_{T_2}$ and $G_\sigma$, but transforms $\phi_R (\vec r)$ as
\beq \label{rotation2}
\phi_R(r_1,r_2,\alpha_s) &\rightarrow& p_1 \pi r_1 r_2 +(p_4+1)\pi r_2 \\
&+& \frac{p_1 \pi}{2}r_2 (r_2 - 1) +\frac{p_3\pi}{2} + \pi + \pi\delta_{\alpha_s,f}\q,\nonumber
\eeq
Here, $\delta_{\alpha_s,f}=1$ when $\alpha_s=f$ and zero otherwise. In contrast to the case of
Kagome lattice where the term $p_4\pi r_2$ can be gauged away by the transformation $G_5$,
it becomes $p_4\pi \delta_{\alpha_s,f}$ in the star lattice. Moreover, we can neglect the constant $\pi$ because it correspond to an IGG operation.
Finally, we arrive at
\beq
\phi_R(r_1,r_2,\alpha_s) &=& p_1 \pi r_1 r_2 + \frac{p_1 \pi}{2}r_2 (r_2 - 1)\\
&+& \frac{p_3 \pi}{2} + p_4 \pi \delta_{\alpha_s,f} .
\eeq


\begin{thebibliography}{99}
\bibitem{zheng} Y.Z. Zheng et. al., Angew. Chem. Int. Ed. {\bf 46}, 6076 (2007).
%\bibitem{coldea} R. Coldea, D.A. Tennant, and Z. Tylczynski, Phys. Rev. B {\bf 68}, 134424 (2003).
\bibitem{shimizu} Y. Shimizu, K. Miyagawa, K. Kanoda, M. Maesato, and G. Saito, Phys. Rev. Lett. {\bf 91}, 107001 (2003).
%\bibitem{masutomi} R. Masutomi, Y. Karaki, and H. Ishimoto, Phys. Rev. Lett. {\bf 92}, 025301 (2004).
\bibitem{helton} J. S. Helton et. al., Phys. Rev. Lett. {\bf 98} 107204 (2007).
\bibitem{hiroi} Z. Hiroi, M. Hanawa, N. Kobayashi, M. Nohara and H. Takagi, J. Phys. Soc. Jpn. {\bf 70}, 3377 (2001).
\bibitem{okamoto-Kagome} Y.Okamoto, H.Yoshida, Z.Hiroi, arXiv:0901.2237.
\bibitem{okamoto} Y. Okamoto, M. Nohara, H. Aruga-Katori, and H. Takagi, Phys. Rev. Lett. {\bf 99}, 137207 (2007).
\bibitem{ran} Y.Ran, M. Hermele, P.A. Lee and X.G. Wen, Phys. Rev. Lett. {\bf 98}, 117205 (2007).
\bibitem{ryu} S. Ryu, O.I. Motrunich, J. Alicea, and M.P.A. Fisher, Phys. Rev. B. {\bf 75}, 184406 (2007).
\bibitem{lawler2008} M.J. Lawler, A. Paramekanti, Y.B. Kim and L. Balents, Phys. Rev. Lett. {\bf 101}, 197202 (2008).
\bibitem{zhou} Y.Zhou, P.A. Lee, T.K. Ng, and F.C. Zhang, Phys. Rev. Lett. {\bf 101}, 197201 (2008).
\bibitem{podolsky} D. Podolsky, A. Paramekanti, Y.B. Kim and T. Senthil, arXiv:0811.2218.
\bibitem{lecheminant} P.Lecheminant, B.Bernu, C. Lhuillier, L.Pierre and P. Sindzingre, Phys. Rev. B {\bf 56}, 2521 (1997).
\bibitem{waldtmann} Ch. Waldtmann, H.U. Everts, B.Bernu, C.Lhuillier, P. Sindzingre, P. Lecheminant and L. Pierre, Eur. Phys. J. B {\bf 2}, 501 (1998).
\bibitem{nikolic} P.Nikolic and T. Senthil, Phys Rev. B {\bf 68}, 214415 (2003).
\bibitem{singh} R.R.P. Singh and D.A. Huse, Phys. Rev. B {\bf 76}, 180407(R) (2007).
\bibitem{yang2008} B.J. Yang, Y.B. Kim, J.Yu, K.Park, Phys. Rev. B {\bf 77}, 224424 (2008).
\bibitem{lawler2007} M.J. Lawler, L.Fritz, Y.B. Kim and S. Sachdev, Phys. Rev. Lett. {\bf 100}, 187201 (2007).
\bibitem{wang2007} F.Wang, A.Vishwanath, and Y.B. Kim, Phys. Rev. B {\bf 76}, 094421 (2007).
\bibitem{q_order} X.G. Wen, Phys. Rev. B {\bf 65}, 165113 (2002).
\bibitem{sachdev-Kagome} S. Sachdev, Phys. Rev. B {\bf 45}, 12377 (1992).
\bibitem{fawang} F. Wang and A. Vishwanath, Phys. Rev. B {\bf 74}, 174423 (2006).
\bibitem{richter} J. Richter, J. Schulenburg, A. Honecker and D. Schmalfu$\beta$, Phys. Rev. B {\bf 70}, 174454 (2004).
\bibitem{misguich} G. Misguich and P. Sindzingre, J. Phys.: Condens. Matter {\bf 19}, 145202 (2007).
\bibitem{fjaerestad} J.O. Fjaerestad, arXiv:0811.3789.
\bibitem{yao} H.Yao and S.A.Kivelson, Phys. Rev. Lett. {\bf 99}, 247203 (2007).
\bibitem{oleg} O. Tchernyshyov, R. Moessner and S.L. Sondhi, Europhys. Lett. {\bf 73}, 278 (2006).
\bibitem{isakov} S.V. Isakov, T. Senthil, Y.B. Kim, Phys. Rev. B {\bf 72}, 174417 (2005).
\bibitem{sachdev89} N. Read and S. Sachdev, Nucl. Phys. B {\bf 216}, 609 (1989).
\bibitem{affleck89} J. B. Marston and I. Affleck, Phys. Rev. B {\bf 39}, 11538 (1989).
\bibitem{marston} J.B. Marston and C.Zeng, J. Appl. Phys. {\bf 69}, 5962 (1991).
\bibitem{VBS} N. Read and S. Sachdev, Phys. Rev. B {\bf 42}, 4568 (1990).
\bibitem{coleman}  An alternative Sp(N) formulation is developed in R. Flint, M. Dzero and P. Coleman, Nature Physics {\bf 4}, 643 (2008). In our work, we follow the formulation in Ref.\onlinecite{sachdev-Kagome} for simplicity.
\bibitem{wen1} X.G. Wen, Phys. Lett. A {\bf 300}, 175 (2002).
\bibitem{shastry_lattice} C.H. Chung, J.B. Marston and S. Sachdev, Phys. Rev. B {\bf 64}, 134407 (2001).
\bibitem{lawler} M.J. Lawler, H.Y. Kee, Y.B. Kim and A. Vishwanath, Phys. Rev. Lett. {\bf 100}, 227201 (2008).
\bibitem{motrunich} O.I. Motrunich, Phys. Rev. B {\bf 72}, 045105 (2005).
\end{thebibliography}
\end{document}